\documentclass[groupedaddress,superscriptaddress,aps,showpacs,amsmath,amssymb,floatfix, prl,twocolumn,a4]{revtex4-1}

 \usepackage{graphicx,float}
 \usepackage{subfigure}
 \usepackage{dcolumn}
\usepackage{bm}
\usepackage{mathrsfs}
\usepackage{txfonts}
\usepackage{CJK}
\usepackage{SIunits}
\usepackage{epsfig}
\usepackage{lipsum}
\usepackage{color}

 \newenvironment{SChinese}{%
  \CJKfamily{gbsn}%
 \CJKtilde
  \CJKnospace}{}

\allowdisplaybreaks[2]

\begin{document}

\begin{CJK}{UTF8}{} 
\begin{SChinese}

\title{Cavity-free nondestructive detection of a single optical photon}

 \author{Keyu Xia (夏可宇)}  %
 \email{keyu.xia@mq.edu.au}
 \affiliation{Centre for Engineered Quantum Systems, Department of Physics and Astronomy, Macquarie University, NSW 2109, Australia}

 \author{Mattias Johnsson}  %
 \affiliation{Centre for Engineered Quantum Systems, Department of Physics and Astronomy, Macquarie University, NSW 2109, Australia}

 \author{Peter L. Knight}
 \affiliation{Department of Physics, Blackett Laboratory, Imperial College, London SW72AZ, UK}

 \author{Jason Twamley}
 \affiliation{Centre for Engineered Quantum Systems, Department of Physics and Astronomy, Macquarie University, NSW 2109, Australia}

\date{\today}

\begin{abstract}
  Detecting a single photon without absorbing it is a long standing challenge in quantum optics. All experiments demonstrating the nondestructive detection of a photon make use of a high quality cavity. We present a cavity-free scheme for nondestructive single-photon detection. By pumping a nonlinear medium we implement an inter-field Rabi-oscillation which leads to a $\sim\pi-$phase shift on weak probe coherent laser field in the presence of a single signal photon without destroying the signal photon. Our  cavity-free scheme operates with a fast intrinsic time scale in comparison with similar cavity-based schemes. We implement a full real-space multimode numerical analysis of the interacting photonic modes and confirm the validity of our nondestructive scheme in the multimode case.
\end{abstract}


\maketitle

\end{SChinese}
 \end{CJK}

Unlike ``demolition" detection \cite{SNW, *APD,SQMW1,*SQMW2}, quantum non-demolition (QND) or ``nondestructive" photon detection ideally avoids any absorption of the photons during the measurement. It is essential for applications in quantum information processing as it can nondestructively monitor the outcome of the system interaction and the operation of quantum gates.

Due to the ultrastrong coupling between a high quality microwave (mw) cavity and a qubit, nondestructive single-photon detection has 
been demonstrated in mw cavity quantum-electrodynamics (QED) \cite{PTomog,QZE,Collapse,ROQND,QGate,Stark1,Stark2}. A  breakthrough result demonstrated  a  $\pi$ phase shift of an  atom resonantly coupled to a high-Q mw cavity after a full-cycle Rabi oscillation (RO) \cite{ROQND}. 
However, the experimental verification of the nondestructive detection of an optical photon in an optical cavity has only been reported very recently \cite{QNDOpthicalPhoton}. So far, all experimental proposals, except for the microwave proposal by  Sathyamoorthy et al. \cite{QNDTravelMW} using a chain of superconducting transmons strongly coupled to a propagating mw photon,  detect a mw or optical single photon by exploiting sophisticated cavity-QED designs. 

Nondestructive detection of a traveling single photon is a key element in quantum information processing but is even more challenging  compared to the detection of a single photon in a cavity. A strong traveling classical light signal field in a Kerr nonlinear medium can cause cross phase modulation (XPM) on a weak copropagating probe field \cite{MilburnXPM,*XPMExp1,QNDReview}. However, the probe field's phase shift induced by a single signal photon is typically too small to be detected. Therefore, the nondestructive detection of a single photon exploiting the cross Kerr nonlinearity has proved to be a particular challenge in the field of quantum optics \cite{QNDReview}. Researchers have suggested approaches that enhance the weak cross-Kerr effect for single-photon phase modulation \cite{XPM} and detection \cite{QNDReview} but its use for nondestructive measurement has been questioned  \cite{KerrDebate1,KerrDebate2,KerrDebate4,*KerrDebate3,*ShapiroXPM} if the ``Ising"-type interaction, $\hat{a}^\dag \hat{a}\hat{b}^\dag \hat{b}$, between the signal and probe modes: $\hat{a}$ and $\hat{b}$, is utilized.
In this letter, we theoretically propose a method for nondestructive detection of a single traveling photon without using a cavity. In our proposal, we apply a tunable three-mode interaction $\hat{a}_a \hat{a}^\dag_p \hat{a}^\dag_s + \hat{a}^\dag_a \hat{a}_p \hat{a}_s$ resulting in ``Rabi"-like oscillation in mode occupation.



\begin{figure}
 \centering  
 \includegraphics[width=0.95\linewidth]{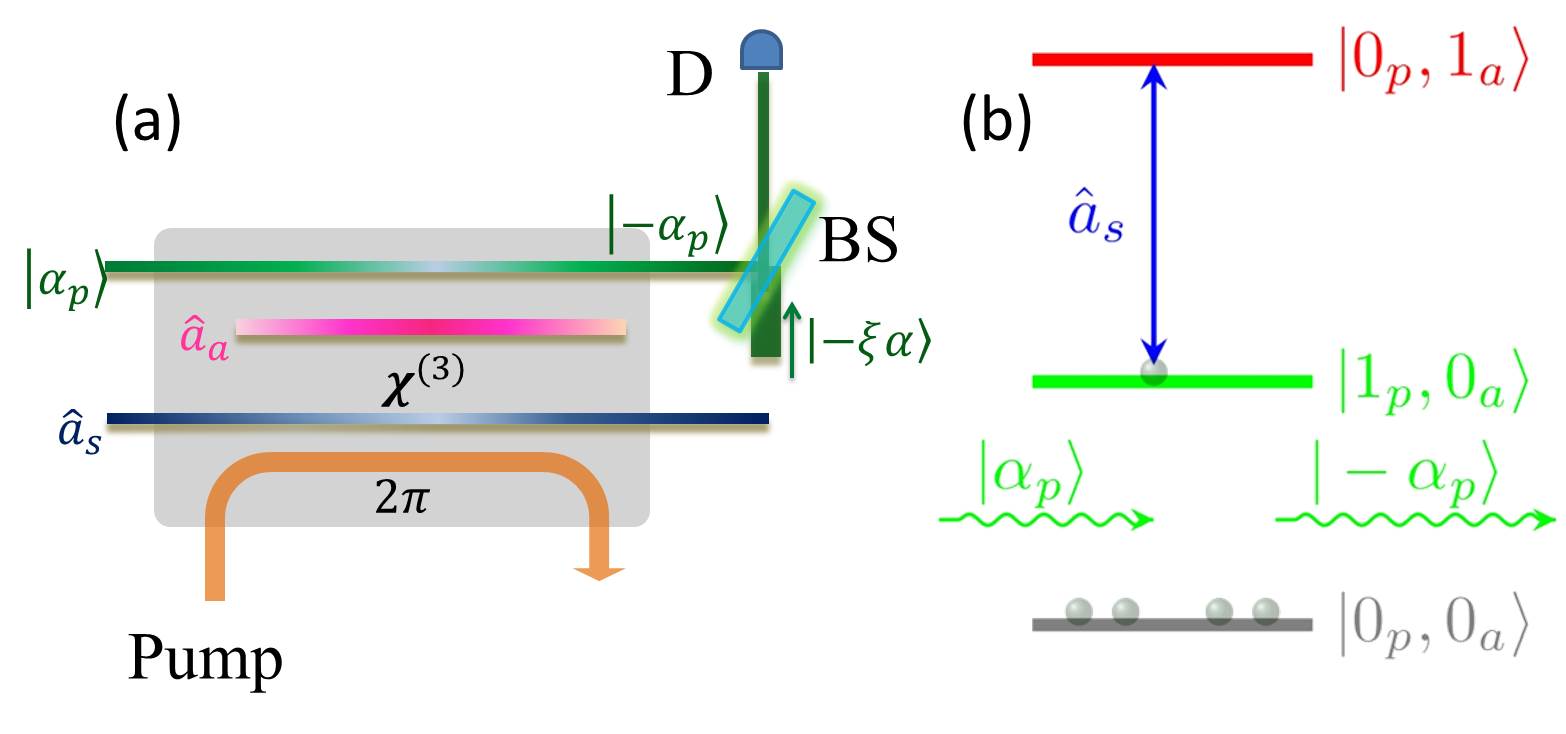}\\
 \caption{(Color online) Schematic for detection of a single traveling photon. (a) configuration for nondestructive detection of a single traveling photon in a  nonlinear medium. A weak probe coherent field $|\alpha_p\rangle$ (green line) interacts with the signal and auxiliary modes $\hat{a}_s$ (blue line) and $\hat{a}_a$ (magenta line) in a third-order nonlinear ($\chi^{(3)}$) medium. 
The probe field, after transmitting through the medium, is displaced via a highly reflective beam splitter and then is detected. 
(b) level diagram describing the interaction between the signal, auxiliary and probe photons.
} \label{fig:system}
\end{figure}

We begin by discussing the main concept for the nondestructive detection of a single photon by describing the setup as depicted in Fig. \ref{fig:system}(a). As in \cite{Nature478p360}, we strongly pump a  nonlinear optical medium to create a standard three-wave-mixing Hamiltonian. In this setup, the nonlinear coupling among the auxiliary mode $\hat{a}_a$, the signal field  $\hat{a}_s$ and the probe mode $\hat{a}_p$ is enhanced and dynamically controlled by the strong pump (control) field $E_c$ with frequency $\omega_p$ and propagation constant $k_p$. The modes $\hat{a}_a$, $\hat{a}_s$ and $\hat{a}_p$ are assumed to have the central frequencies $\omega_a$, $\omega_s$ and $\omega_p$,  propagation constants $k_a$, $k_s$ and $k_p$, and  durations $\tau_a$, $\tau_s$ and $\tau_p$, respectively. We assume that the three modes  suffer zero dispersion over their individual bandwidths and propagate with constant group velocities. This is reasonable if the duration of the signal and probe fields are long enough to span a narrow bandwidth. 
By properly choosing the length $L$ of the nonlinear medium and controlling the intensity of the pumping, we can enable an accumulated interaction of $2\pi$ among the three fields $\hat{a}_a$, $\hat{a}_s$ and $\hat{a}_p$ as they propagate through  length $L$ of the medium.
The interaction Hamiltonian describing the three-wave mixing among the fields in the medium (along the $z$ direction) $\hat{a}_a$, $\hat{a}_s$ and $\hat{a}_p$ takes the form $(\hbar=1)$
\begin{equation}\label{eq:HI}
 \hat{H}_I = \frac{g(E_c)}{2} \hat{a}_a \hat{a}^\dag_p \hat{a}^\dag_s +  \frac{g^*(E_c)}{2} \hat{a}^\dag_a \hat{a}_p \hat{a}_s \;,
\end{equation}
where $g(E_c)$ indicates the nonlinear coupling strength which can be tuned by the intensity of the pump field $E_c$.

Next we explain how to induce a substantial phase shift in the weak probe field to distinguish between the vacuum and single-photon states of the signal field.
We arrange that the auxiliary field is initially in the vacuum state $|0\rangle_a$. The signal field can be either the vacuum state $|0\rangle_s$ or a single-photon state $|1\rangle_s$. The weak probe coherent field $|\alpha_p\rangle$ copropagates with the signal mode and the pumping field along the $z$ direction. We assume that $\alpha_p \ll 1$ such that $|\alpha_p\rangle \approx |0\rangle_p + \alpha_p |1\rangle_p$. We schematically depict the interaction between the photons  by the level diagram shown in Fig. \ref{fig:system}. For a weak probe field $\alpha_p \ll 1$, the joint quantum state $|0_p,0_a\rangle$ is mostly populated, while the occupation of the state $|1_p,0_a\rangle$ is approximately $|\alpha_p|^2\exp(-|\alpha_p|^2/2)$. The signal field drives the transition $|1_p,0_a\rangle \leftrightarrow |0_p,1_a\rangle$. To provide a simple but transparent picture for our idea we consider the ideal case of $\omega_c + \omega_a =\omega_p + \omega_s$ and $k_c + k_a =k_p + k_s$. If $g(E_p) L=2\pi$ and a single photon in the signal field, i.e. $\hat{a}_s^\dag |\O{}\rangle = |1\rangle_s$, (where $|\O{}\rangle$ is the vacuum), the signal field will drive the transition $|1_p,0_a\rangle \rightarrow |0_p,1_a\rangle$ and then bring back the occupation to $|1_p,0_a\rangle$. In this case, the state $|1_p,0_a\rangle$ suffers a $\pi$ phase shift and therefore the probe field becomes $|\Psi\rangle_t=|0_p\rangle - \alpha_p |1\rangle \approx |-\alpha_p\rangle$. For a vacuum signal field $|0_s\rangle$, the probe field has no phase shift after propagating through the medium. As a result, the transmitted field is $|\Psi\rangle_t=|\alpha_p\rangle$. In our numerical simulation, the transitions $|(n+1)_p,0_a\rangle \rightarrow |n_p,1_a\rangle$ involving higher Fock states with $n_p>0$ are also taken into account by truncating the probe field at a large Fock state.

To detect the change in the probe field, we displace the probe field after it has propagated/transmitted through the nonlinear medium by a coherent field $|-\alpha_p\rangle$. Then the density matrix of the detected field is related to that of the transmitted field by $\rho_D(-\alpha_p) = D(-\alpha_p)\rho_t D(\alpha_p)$ \cite{HarocheDispl}, where $D(\alpha_p)$ is the displacement operator. To do so, we inject a large coherent field $|-\xi\alpha_p\rangle$ ($\xi \gg 1$), in to a highly reflective beam splitter ($(1-\eta) : \eta$) with a reflectivity of $\eta>0.99$, together with the probe field, so that the transmission is $|-\alpha_p\rangle$ \cite{BS1}. In doing so, the detector will measure a coherent field $|\Psi\rangle_D=|-2\alpha_p\rangle$ if a single photon is present, but no photon ($|\Psi\rangle_D=|0_p\rangle$ ) is detected if no signal photon is input into the system. 

We next evaluate the performance of our single-photon detection. We first treat the propagation of the optical fields via a single mode approach governed by a quantum Langevin equation by replacing the time $t$ with the position $z$
\begin{equation}\label{eq:MEq}
 \partial \rho/\partial z = -i [\hat{H}_I,\rho] + \mathcal{L}\rho \;, 
\end{equation}
where $\rho$ is the density matrix of the system at the position $z$, and $\mathcal{L}\rho$ takes into account the possible decoherence, $\mathcal{L}\rho=\sum_{j=a,p,s} \frac{\gamma_j}{2}\left(2\hat{a}_j \rho\hat{a}^\dag_j  -\hat{a}^\dag_j \hat{a}_j \rho -\rho \hat{a}^\dag_j \hat{a}_j\right)$ with $\gamma_j$ is the loss per meter of the $j$th mode. Numerically solving the quantum Langevin equation, one can calculate the transmitted state of the probe field and the detected state. As a test, the Langevin equation reproduces all the results shown in Fig. 2(a) of the work \cite{Nature478p360}. We will implement a more thorough multi-mode propagation analysis later on but for now we take $\gamma_a=\gamma_p=\gamma_s=10^{-3} ~\text{rad}\cdot \text{m}^{-1}$.

We consider the click at the detector as evidence of a single photon in the signal mode. If no click is observed, then no photon is in the signal mode. We assume perfect quantum efficiency of the detector. Thus the vacuum state of the signal mode yields no click on the detector with $100\%$ probability. When a single photon is injected into the system, the state presented to the detector is $|\Psi\rangle_D=|-2\alpha_p\rangle$. This state can generate a click with a certain probability but the small vacuum element in $|\Psi\rangle_D$ can lead to a detection failure of the signal photon with a probability $P_\text{err}$. Therefore, taking $\alpha_p\in \Re$, this error probability is \cite{StatePerr} given by $P_\text{err} (\alpha_p)  = \text{Tr}[|0_p\rangle \langle 0_p| D(-\alpha_p) \rho(L) D(\alpha_p)] 
\approx |\langle 0_p| -2\alpha_p\rangle|^2= e^{-4|\alpha_p|^2}$.
We note that this error decreases exponentially with $\alpha_p$ but we will see below that the non-destructiveness of the protocol also degraded with increasing $\alpha_p$. 
If we explore an N-cascade detection configuration \cite{QNDTravelMW,Nature478p360}, and consider a single click on any detector as an indicator for a single photon in the signal mode, then the error probability of failing to detect the signal photon decreases as $P^N_\text{err}$. This error decreases rapidly to vanishing small values as $N$ increases. 

Another parameter indicating the ``non-destructive'' performance of the detection is the fidelity of the signal single-photon state $\mathcal{F}(\alpha_p,L)=\text{Tr}[\rho(\alpha_p,L) |1_s\rangle \langle 1_s|]$, after detection. The value of $\mathcal{F}$ can be evaluated numerically and for low values of $\alpha_p$ we find it close to unity.

Next we estimate the performance of our system by numerically simulating the {\em mode} master equation Eq. \ref{eq:MEq}.
For a vacuum signal state $|0_s\rangle$, the detected field after the beam splitter is trivial, being the initial probe vacuume field $|0_p\rangle$.  Thus we are primarily interested the detection field $|\Psi\rangle_D$ in the case of a single-photon signal input $|1_s\rangle$. 
\begin{figure}
 \centering
\includegraphics[width=0.99\linewidth]{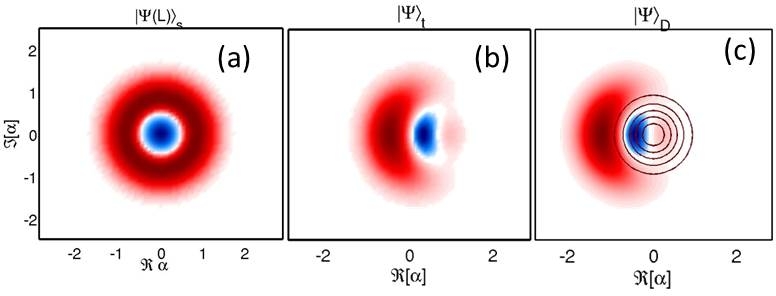}\\
 \caption{(Color online) Wigner functions of the transmitted and detected states for a probe field with $|\alpha_p|^2=0.6$. (a) Transmitted signal state $|\Psi\rangle_s$ after interacting for the length of the media; (b) Transmitted probe field $|\Psi\rangle_t$ after interacting for the length of the media; (c) Detected state $|\Psi\rangle_D$ presented to detector. The concentric circles show the Wigner function contours of the detection field in the case of an input signal vacuum state.}\label{fig:Wignerfun}
\end{figure}
First, for a certain probe field of $|\alpha_p|^2=0.6$, and a single-photon signal field $|1\rangle_s$, we investigate to what extent the phase of the transmitted probe field can be shifted while the signal field remains unchanged. Figure \ref{fig:Wignerfun} shows the Wigner functions of the transmitted and detected field. In Fig. \ref{fig:Wignerfun}(a), the Wigner function indicates the transmitted signal field at $z=2\pi$ is almost a single photon state  $|1\rangle_s$. The fidelity is evaluated to be $\mathcal{F}\approx 0.9$. The transmitted probe field $|\Psi\rangle_t$ has a complicated Wigner function which includes a region with a large negative value. It means that $|\Psi\rangle_t$ is a quantum field at $z=2\pi$ because the initial probe field includes considerable occupation in higher Fock states $|n_p\rangle$ with $n_p>1$. As a result, the fidelity $|\langle -\alpha_p |\Psi\rangle_t|^2$ is about $0.73$, see Fig. \ref{fig:Wignerfun}(b). After displacing by $|-\alpha_p\rangle$ with the beam splitter, the Wigner function (Fig. \ref{fig:Wignerfun}(c)) moves to the left by $\alpha_p$ with the result that the detected field $|\Psi\rangle_D$ can be well distinguished from the field of $|0_p\rangle$, the latter being the detected field  in the case of a vacuum input signal.

%
During the time evolution of three modes, the occupation of the probe field displays Rabi-like oscillation as the fields propagates through the medium. The occupation of the probe field first transfers to the auxiliary mode and then returns to the probe field at $gz=2\pi$. The fidelity of the signal mode also oscillates with the propagation distance $z$. This process repeats every $gz=2\pi$. It is found that, given a small $|\alpha_p|^2$, the phase of the probe field is shifted by $\pi$ at $gz=2\pi$, but the loss of the signal mode is negligible. For example, when $|\alpha_p|^2=0.6$, the probe field acquires a phase shift of $\sim\pi$ at $gz=2\pi$, while an occupation of $0.894$ in the signal mode yields a high fidelity for non-demolition $\mathcal{F}\approx 0.9$. As $|\alpha_p|^2$ increases the fidelity of the signal mode decreases gradually. If $|\alpha_p|^2$ increases to $0.8$, the non-demolition fidelity $\mathcal{F}$ drops to $0.84$. The signal mode is more greatly modified  by a stronger probe field because the more populated higher Fock states with $n_p>1$ interact dispersively with the signal mode \cite{FockDispersion}. 
 As a result the transitions involving higher Fock states prevent us from perfect non-demolition of the signal mode.

A ``good" single-photon detection requires a high fidelity $\mathcal{F}$ of the signal mode and a small detection error $P_\text{err}$.
Reducing the mean photon number of the probe field can increase the fidelity of the transmitted signal mode but  also leads to an increase in the detection error because it provides less signal to the detector.
\begin{figure}
 \centering
 \includegraphics[width=0.5\linewidth]{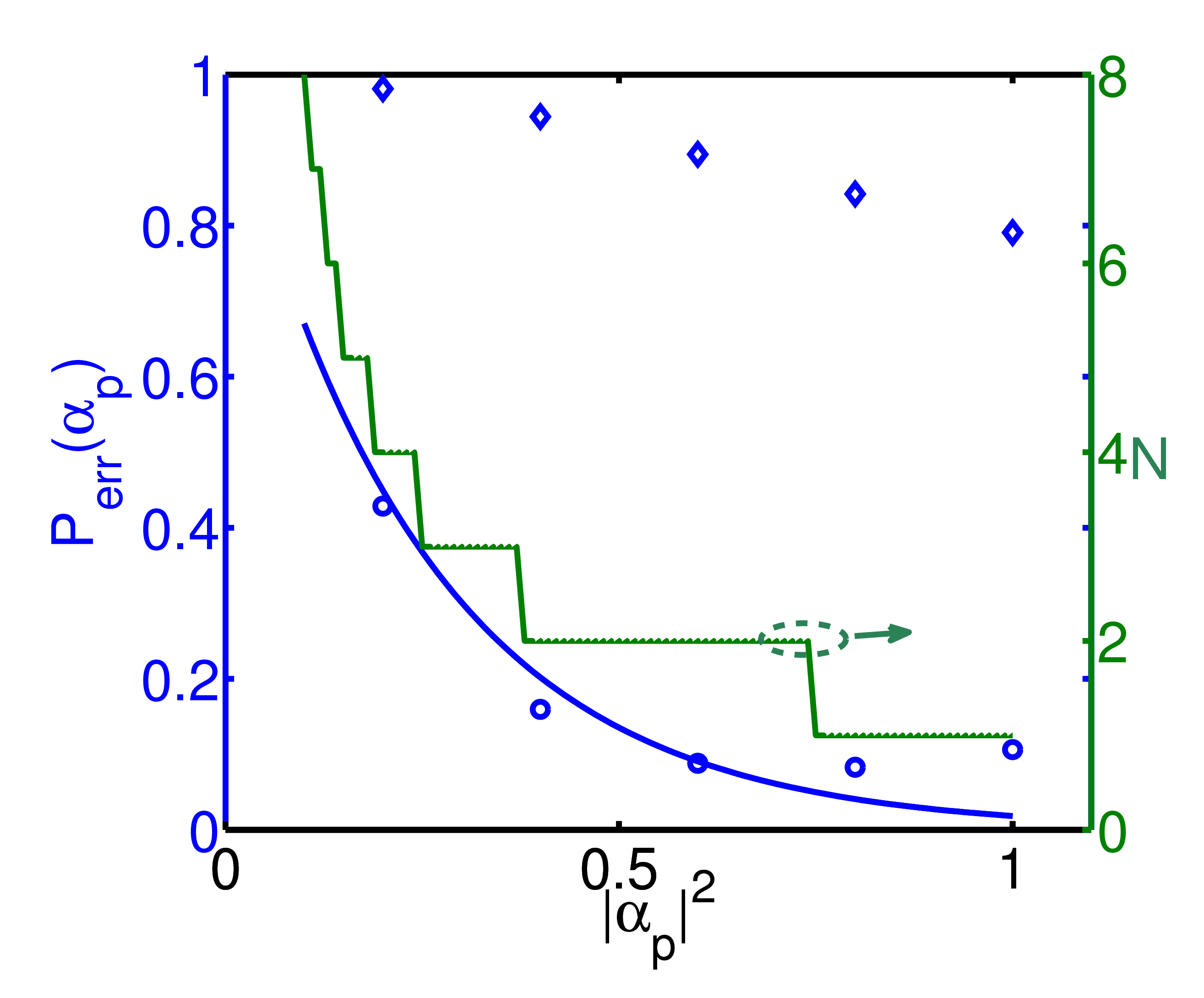}\\
 \caption{(Color online) Detection error, $P_{err}$, and fidelity $\mathcal{F}$, as a function of the probe field. Blue line indicates the analytic estimation. Circles show the numerical data for different probe fields, while the diamond data (with regard to left vertical axis) shows the fidelity of the transmitted signal mode. The green line (right vertical axis), gives the number $N$ of cascaded detection units to achieve an overall detection error $P^N_\text{err}<5\%$.} \label{fig:Perr}
\end{figure}
 For one measurement, the error probability $P_\text{err}\approx e^{-4|\alpha_p|^2}$ decreases exponentially as $|\alpha_p|^2$ increases. From Fig.\ref{fig:Perr}, this analytic form (blue line) is in good agreement with the numerical results (circles) for small $|\alpha_p|^2$. The numerical value of $P_\text{err}$ reaches its minimum $0.09$ at $|\alpha_p|^2=0.6$ and then increases slowly as $|\alpha_p|^2$ increases. On the other hand, the fidelity $\mathcal{F}$ decreases as well. When $|\alpha_p|^2=0.6$, $\mathcal{F}=0.9$. An efficient QND detection requires $\mathcal{F}/P_\text{err} \gg 1$. Thus, a probe field of $|\alpha_p|^2 \sim 0.6$ is optimal and yields $\mathcal{F}/P_\text{err}=10$. A N-cascade configuration of the detection unit can reduce the error possibility to $P^N_\text{err}$ and relax the requirement $\mathcal{F}^N/P^N_\text{err} \gg 1$. The green line in Fig.\ref{fig:Perr} provides an estimate of the $N$ required to achieve $P^N_\text{err}<5\%$. Even for a very weak probe field of $|\alpha_p|^2 =0.2$, four cascaded detection units can achieve $P^4_\text{err}\sim 3\%$ and $\mathcal{F}^4/P^4_\text{err} >27$.

Although QND measurement of photons using XPM in a Kerr nonlinear medium has been proposed for decades \cite{QNDReview,XPM}, it has been widely doubted \cite{KerrDebate1,KerrDebate2,KerrDebate4,*KerrDebate3,*ShapiroXPM} when a continuous-time multi-mode model \cite{KerrDebate2}, or a finite response time \cite{KerrDebate4,*KerrDebate3,*ShapiroXPM}, is considered and has never been demonstrated experimentally at the single-photon level. However our scheme exploiting ``Rabi"-like oscillation overcomes most of these difficulties. Above we  discuss the performance of our nondestructive single-photon detection using four-wave mixing in a single-mode model. In what follows, we investigate the validity of our scheme by numerically simulating the interaction of continuous-time multi-mode quantum wave packets propagating in real space.

For our purpose of single-photon QND detection, we only need to address the fidelity and phase shift of a photon-pair input state $|1_p,1_s\rangle$ after propagating a certain distance.
During the propagation of the probe and signal fields the auxiliary field can be excited. To simulate the interaction we define an associated wave function  $\phi_{ps}(t;z_p,z_s)$ for the state $|1_p,1_s\rangle$, and the wave function $\phi_{a}(t;z_a)$ for the auxiliary field state $|1_a\rangle$. These wave functions imply that the photons $|1_p\rangle$ and $|1_s\rangle$ ($|1_a\rangle$) appear(s) at $z_p$ and $z_s$ ($z_a$)  at time $t$ with probability of $|\phi_{ps}(t;z_p,z_s)|^2$ ($|\phi_{a}(t;z_a)|^2$). The state of the fields can be described by a general wave function
 $|\phi(t;z_a,z_p,z_s)\rangle =  \int dz_a \phi_a(t;z_a) \hat{a}_a^\dag |\O{}\rangle 
  + \int\int dz_pdz_s \phi_{ps}(t;z_p,z_{s}) \hat{a}_p^\dag \hat{a}_s^\dag|\O{}\rangle$,
where $|\O{}\rangle$ is the vacuum state.  We apply a Gaussian input $\phi_{ps}(t;z_{p},z_s)=\frac{1}{\sqrt{\pi\tau_p\tau_s}} e^{-(z_p-z_{p,0})^2/2\tau_p^2} e^{-(z_s-z_{s,0})^2/2\tau_s^2}$,  where $z_{p,0}$ and $z_{s,0}$ are the group delays of the probe and signal wave functions, respectively. For simplicity, we assume $\tau_p =\tau_s=\tau$. Assuming this product input yields no quantum (or classical) correlations in the input we have initially, $\int\int dz_p dz_s |\phi_{ps}(0;z_p,z_s)|^2=1$ and $\int dz_a |\phi_{a}(0;z_a)|^2=0$.  The evolution of the photonic wave functions is governed by the partial differential equations (PDEs) \cite{SinglePhotonRealSpace1,*SinglePhotonRealSpace5,*SinglePhotonRealSpace6,*SinglePhotonRealSpace2,*SinglePhotonRealSpace3,*SinglePhotonRealSpace4}
\begin{equation}\nonumber
 \begin{split}
\frac{\partial \phi_{ps}}{\partial t} & = -v_p \frac{\partial \phi_{ps}}{\partial z_p} -v_s \frac{\partial \phi_{ps}}{\partial z_s} - \frac{ig_0}{2} \int_0^L f_g(z_a,z_p,z_s)  \phi_{a}dz_a  \;,\\
 \frac{\partial \phi_{a}}{\partial t} & = -v_a \frac{\partial \phi_{a}}{\partial z_a}  - \frac{ig_0}{2} \int_0^L\int_0^L f_g(z_a,z_p,z_s)  \phi_{ps}dz_pdz_s\;,
 \end{split} \label{eqn4}
\end{equation}
where we assume perfect phase and energy matching $\Delta k = (k_c+k_a-k_p-k_s)=0$ and $\Delta_a = (\omega_c+\omega_a-\omega_p-\omega_s)=0$, respectively. The nonlinear medium is assumed to possess a spatial nonlocal response distribution \cite{NonlocalKerr1,*NonlocalKerr2,*NonlocalKerr3,*NonlocalKerr4,*NonlocalKerr5,*NonlocalKerr6} $f_g(z_a,z_p,z_s) = \frac{1}{\sqrt{\pi \sigma^3}}e^{-\frac{(z_a-z_p)^2}{2\sigma^2}}e^{-\frac{(z_a-z_s)^2}{2\sigma^2}}$, where $\sigma$ indicates the finite interaction length. The prefactor of this spatial response function is not important for experimental implementation because the coupling strength can be tuned via the pump laser power. 
 The photon pulses  are long enough to assume that the group velocity of each mode is constant in time (we assume $v_a=v_p=1$ ). The fidelity of the photon pair state $|1_p,1_s\rangle$ is evaluated as 
 $\mathcal{F}=\frac{1}{2}|1-\int \int dz_p dz_s \phi^*_\text{ps,in}(t_\text{e};z_{p},z_s;z_\text{p,e},z_\text{s,e})\phi_{ps}(t_\text{e};z_{p},z_s;z_\text{p,e},z_\text{s,e})|$,
 where $\phi^*_\text{in}(t_\text{e};z_{p},z_s;z_\text{p,e},z_\text{s,e})$
 means the input wave packets $\phi_{ps}(t=0)$ freely propagating to the output position $z_\text{p,e}$ and $z_\text{s,e}$. With this definition  we expect the fidelity to be $\mathcal{F}=0$ when no phase shift is present but $\mathcal{F}=1$ for a $\pi$ phase shift. Throughout our simulation, we set $\sigma=0.2$ and the duration $\tau=0.6$ in $z$.

\begin{figure}
 \centering 
 \includegraphics[width=0.96\linewidth]{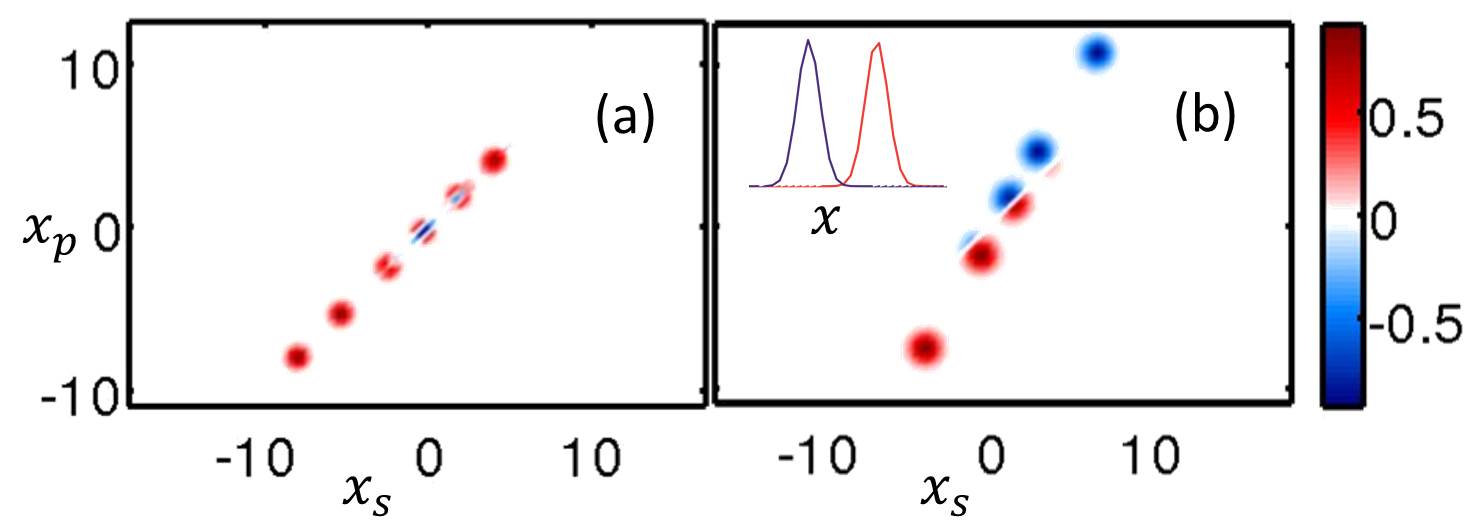}\\
 \caption{(Color online) Evolution of the wave function $\phi_\text{ps}$ for (a) the same propagating speeds $v_\text{p}=v_\text{s}=1$ and delay, and (b) different speeds $v_\text{p}> v_\text{s}$ ($v_\text{p}=1$ but $v_\text{s}=0.6$) and different delays, $x_{s0}=-4$ and $x_{p0}=-7.6$. Inset plot in Fig. (b) shows the positions of input probe (blue line) and signal (red line) wave functions corresponding to the first red spot. 
}\label{fig:Evolphips}
\end{figure}
We numerically simulate the time evolution of the associated wave function $\phi_\text{ps}$ at different positions in real space using the xmds package to solve the PDEs \cite{xmds}. Here the probe and signal wave functions initially have no  correlation yielding a form for  $\phi_\text{ps}$ of a 2D symmetric Gaussian. In Fig. \ref{fig:Evolphips}(a), the probe and signal wave functions have the same delay and are assume to copropagate at same group velocity, $v_\text{p}=v_\text{s}=1$. We find that only the diagonal region of the wave function $\phi_{ps}$ has a $\pi$ phase shift because the nonlinear interaction occurs only when the probe and signal photons overlap in space, $|z_p - z_s| < \sigma$. In Fig. \ref{fig:Evolphips}(b), the signal pulse enters the medium first and is followed by the probe pulse (see inset plot). The probe mode propagates faster than the signal mode. This configuration has been used for XPM to induce a large phase shift using an Ising type interaction but the resulting fidelity was poor \cite{ShapiroScanOver}. In our case, during the propagation the probe field  scans over the whole signal pulse and subsequently the two fields   interact completely with each other. It is found that the wave function starts to invert when the leading edge of the probe pulse meets the tail of the signal pulse. As a result, a $\pi$-phase shift can be induced after the probe pulse passes through the entire signal pulse and we find that the fidelity is very high, with $\mathcal{F}\approx 1$. 

The experimental implementation of our scheme can be conducted in various systems. We follow the configuration in \cite{Nature478p360}. We use a polarization-maintaining photonic crystal fiber (PCF) as the nonlinear medium and pump it with a $100~\text{\milli\watt} - 1~ \text{\watt}$ cw laser beam at $532 ~\text{\nano\meter}$. By controlling the power of the pump laser and the length $L$ of the PCF, we can achieve a full Rabi oscillation for $L\sim 66$m. We choose a weak $504~\text{\nano\meter}$ long laser pulse with the input state $|\alpha_p\rangle$ as the probe mode $\hat{a}_p$. The $710~\text{\nano\meter}$ as the auxiliary mode is absent in the input. In this setup, we are able to nondestructively detect a single photon at $766~\text{\nano\meter}$. Note that only the pump laser is strong. As a result, this pump laser will modulate the energy of other three modes in the same way and can be neglected. The relative large self modulation of the pump laser energy can also be canceled by tuning the pump wavelength a little. We also note that a slot-waveguide configurations can yield four-wave mixing strengths almost four orders larger than in PCF \cite{trita2011}.

In conclusion, we have proposed a cavity-free scheme to detect a single optical photon without destroying it. In our setup we have induced a $\pi$ phase shift in a weak probe field by a traveling single photon in a nonlinear medium with this single photon remaining largely unchanged. 


\begin{thebibliography}{42}%
\makeatletter
\providecommand \@ifxundefined [1]{%
 \@ifx{#1\undefined}
}%
\providecommand \@ifnum [1]{%
 \ifnum #1\expandafter \@firstoftwo
 \else \expandafter \@secondoftwo
 \fi
}%
\providecommand \@ifx [1]{%
 \ifx #1\expandafter \@firstoftwo
 \else \expandafter \@secondoftwo
 \fi
}%
\providecommand \natexlab [1]{#1}%
\providecommand \enquote  [1]{``#1''}%
\providecommand \bibnamefont  [1]{#1}%
\providecommand \bibfnamefont [1]{#1}%
\providecommand \citenamefont [1]{#1}%
\providecommand \href@noop [0]{\@secondoftwo}%
\providecommand \href [0]{\begingroup \@sanitize@url \@href}%
\providecommand \@href[1]{\@@startlink{#1}\@@href}%
\providecommand \@@href[1]{\endgroup#1\@@endlink}%
\providecommand \@sanitize@url [0]{\catcode `\\12\catcode `\$12\catcode
  `\&12\catcode `\#12\catcode `\^12\catcode `\_12\catcode `\%12\relax}%
\providecommand \@@startlink[1]{}%
\providecommand \@@endlink[0]{}%
\providecommand \url  [0]{\begingroup\@sanitize@url \@url }%
\providecommand \@url [1]{\endgroup\@href {#1}{\urlprefix }}%
\providecommand \urlprefix  [0]{URL }%
\providecommand \Eprint [0]{\href }%
\providecommand \doibase [0]{http://dx.doi.org/}%
\providecommand \selectlanguage [0]{\@gobble}%
\providecommand \bibinfo  [0]{\@secondoftwo}%
\providecommand \bibfield  [0]{\@secondoftwo}%
\providecommand \translation [1]{[#1]}%
\providecommand \BibitemOpen [0]{}%
\providecommand \bibitemStop [0]{}%
\providecommand \bibitemNoStop [0]{.\EOS\space}%
\providecommand \EOS [0]{\spacefactor3000\relax}%
\providecommand \BibitemShut  [1]{\csname bibitem#1\endcsname}%
\let\auto@bib@innerbib\@empty
\bibitem [{\citenamefont {Divochiy}\ \emph {et~al.}(2008)\citenamefont
  {Divochiy}, \citenamefont {Marsili}, \citenamefont {Bitauld}, \citenamefont
  {Gaggero}, \citenamefont {Leoni}, \citenamefont {Mattioli}, \citenamefont
  {Korneev}, \citenamefont {Seleznev}, \citenamefont {Kaurova}, \citenamefont
  {Minaeva}, \citenamefont {Gol'tsman}, \citenamefont {Lagoudakis},
  \citenamefont {Benkhaoul}, \citenamefont {L\'{e}vy},\ and\ \citenamefont
  {Fiore}}]{SNW}%
  \BibitemOpen
  \bibfield  {author} {\bibinfo {author} {\bibfnamefont {A.}~\bibnamefont
  {Divochiy}}, \bibinfo {author} {\bibfnamefont {F.}~\bibnamefont {Marsili}},
  \bibinfo {author} {\bibfnamefont {D.}~\bibnamefont {Bitauld}}, \bibinfo
  {author} {\bibfnamefont {A.}~\bibnamefont {Gaggero}}, \bibinfo {author}
  {\bibfnamefont {R.}~\bibnamefont {Leoni}}, \bibinfo {author} {\bibfnamefont
  {F.}~\bibnamefont {Mattioli}}, \bibinfo {author} {\bibfnamefont
  {A.}~\bibnamefont {Korneev}}, \bibinfo {author} {\bibfnamefont
  {V.}~\bibnamefont {Seleznev}}, \bibinfo {author} {\bibfnamefont
  {N.}~\bibnamefont {Kaurova}}, \bibinfo {author} {\bibfnamefont
  {O.}~\bibnamefont {Minaeva}}, \bibinfo {author} {\bibfnamefont
  {G.}~\bibnamefont {Gol'tsman}}, \bibinfo {author} {\bibfnamefont {K.~G.}\
  \bibnamefont {Lagoudakis}}, \bibinfo {author} {\bibfnamefont
  {M.}~\bibnamefont {Benkhaoul}}, \bibinfo {author} {\bibfnamefont
  {F.}~\bibnamefont {L\'{e}vy}}, \ and\ \bibinfo {author} {\bibfnamefont
  {A.}~\bibnamefont {Fiore}},\ }\href@noop {} {\bibfield  {journal} {\bibinfo
  {journal} {Nature Photon.}\ }\textbf {\bibinfo {volume} {2}},\ \bibinfo
  {pages} {302} (\bibinfo {year} {2008})}\BibitemShut {NoStop}%
\bibitem [{\citenamefont {Kardyna\l}\ \emph {et~al.}(2008)\citenamefont
  {Kardyna\l}, \citenamefont {Yuan},\ and\ \citenamefont {Shields}}]{APD}%
  \BibitemOpen
  \bibfield  {author} {\bibinfo {author} {\bibfnamefont {B.~E.}\ \bibnamefont
  {Kardyna\l}}, \bibinfo {author} {\bibfnamefont {Z.~L.}\ \bibnamefont {Yuan}},
  \ and\ \bibinfo {author} {\bibfnamefont {A.~J.}\ \bibnamefont {Shields}},\
  }\href@noop {} {\bibfield  {journal} {\bibinfo  {journal} {Nature Photon.}\
  }\textbf {\bibinfo {volume} {2}},\ \bibinfo {pages} {425} (\bibinfo {year}
  {2008})}\BibitemShut {NoStop}%
\bibitem [{\citenamefont {Chen}\ \emph {et~al.}(2011)\citenamefont {Chen},
  \citenamefont {Hover}, \citenamefont {Sendelbach}, \citenamefont {Maurer},
  \citenamefont {Merkel}, \citenamefont {Pritchett}, \citenamefont {Wilhelm},\
  and\ \citenamefont {McDermott}}]{SQMW1}%
  \BibitemOpen
  \bibfield  {author} {\bibinfo {author} {\bibfnamefont {Y.-F.}\ \bibnamefont
  {Chen}}, \bibinfo {author} {\bibfnamefont {D.}~\bibnamefont {Hover}},
  \bibinfo {author} {\bibfnamefont {S.}~\bibnamefont {Sendelbach}}, \bibinfo
  {author} {\bibfnamefont {L.}~\bibnamefont {Maurer}}, \bibinfo {author}
  {\bibfnamefont {S.~T.}\ \bibnamefont {Merkel}}, \bibinfo {author}
  {\bibfnamefont {E.~J.}\ \bibnamefont {Pritchett}}, \bibinfo {author}
  {\bibfnamefont {F.~K.}\ \bibnamefont {Wilhelm}}, \ and\ \bibinfo {author}
  {\bibfnamefont {R.}~\bibnamefont {McDermott}},\ }\href@noop {} {\bibfield
  {journal} {\bibinfo  {journal} {Phys. Rev. Lett.}\ }\textbf {\bibinfo
  {volume} {107}},\ \bibinfo {pages} {217401} (\bibinfo {year}
  {2011})}\BibitemShut {NoStop}%
\bibitem [{\citenamefont {Romero.}\ \emph {et~al.}(2009)\citenamefont
  {Romero.}, \citenamefont {Garc\'{i}a-Ripoll},\ and\ \citenamefont
  {Solano.}}]{SQMW2}%
  \BibitemOpen
  \bibfield  {author} {\bibinfo {author} {\bibfnamefont {G.}~\bibnamefont
  {Romero.}}, \bibinfo {author} {\bibfnamefont {J.~J.}\ \bibnamefont
  {Garc\'{i}a-Ripoll}}, \ and\ \bibinfo {author} {\bibfnamefont
  {E.}~\bibnamefont {Solano.}},\ }\href@noop {} {\bibfield  {journal} {\bibinfo
   {journal} {Phys. Rev. Lett.}\ }\textbf {\bibinfo {volume} {102}},\ \bibinfo
  {pages} {173602} (\bibinfo {year} {2009})}\BibitemShut {NoStop}%
\bibitem [{\citenamefont {Brune}\ \emph {et~al.}(2008)\citenamefont {Brune},
  \citenamefont {Bernu}, \citenamefont {Guerlin}, \citenamefont {Del\'eglise},
  \citenamefont {Sayrin}, \citenamefont {Gleyzes}, \citenamefont {Kuhr},
  \citenamefont {Dotsenko}, \citenamefont {Raimond},\ and\ \citenamefont
  {Haroche}}]{PTomog}%
  \BibitemOpen
  \bibfield  {author} {\bibinfo {author} {\bibfnamefont {M.}~\bibnamefont
  {Brune}}, \bibinfo {author} {\bibfnamefont {J.}~\bibnamefont {Bernu}},
  \bibinfo {author} {\bibfnamefont {C.}~\bibnamefont {Guerlin}}, \bibinfo
  {author} {\bibfnamefont {S.}~\bibnamefont {Del\'eglise}}, \bibinfo {author}
  {\bibfnamefont {C.}~\bibnamefont {Sayrin}}, \bibinfo {author} {\bibfnamefont
  {S.}~\bibnamefont {Gleyzes}}, \bibinfo {author} {\bibfnamefont
  {S.}~\bibnamefont {Kuhr}}, \bibinfo {author} {\bibfnamefont {I.}~\bibnamefont
  {Dotsenko}}, \bibinfo {author} {\bibfnamefont {J.~M.}\ \bibnamefont
  {Raimond}}, \ and\ \bibinfo {author} {\bibfnamefont {S.}~\bibnamefont
  {Haroche}},\ }\href@noop {} {\bibfield  {journal} {\bibinfo  {journal} {Phys.
  Rev. Lett.}\ }\textbf {\bibinfo {volume} {101}},\ \bibinfo {pages} {240402}
  (\bibinfo {year} {2008})}\BibitemShut {NoStop}%
\bibitem [{\citenamefont {Bernu}\ \emph {et~al.}(2008)\citenamefont {Bernu},
  \citenamefont {Del\'eglise}, \citenamefont {Sayrin}, \citenamefont {Kuhr},
  \citenamefont {Dotsenko}, \citenamefont {Brune}, \citenamefont {Raimond},\
  and\ \citenamefont {Haroche}}]{QZE}%
  \BibitemOpen
  \bibfield  {author} {\bibinfo {author} {\bibfnamefont {J.}~\bibnamefont
  {Bernu}}, \bibinfo {author} {\bibfnamefont {S.}~\bibnamefont {Del\'eglise}},
  \bibinfo {author} {\bibfnamefont {C.}~\bibnamefont {Sayrin}}, \bibinfo
  {author} {\bibfnamefont {S.}~\bibnamefont {Kuhr}}, \bibinfo {author}
  {\bibfnamefont {I.}~\bibnamefont {Dotsenko}}, \bibinfo {author}
  {\bibfnamefont {M.}~\bibnamefont {Brune}}, \bibinfo {author} {\bibfnamefont
  {J.~M.}\ \bibnamefont {Raimond}}, \ and\ \bibinfo {author} {\bibfnamefont
  {S.}~\bibnamefont {Haroche}},\ }\href@noop {} {\bibfield  {journal} {\bibinfo
   {journal} {Phys. Rev. Lett.}\ }\textbf {\bibinfo {volume} {101}},\ \bibinfo
  {pages} {180402} (\bibinfo {year} {2008})}\BibitemShut {NoStop}%
\bibitem [{\citenamefont {Guerlin}\ \emph {et~al.}(2007)\citenamefont
  {Guerlin}, \citenamefont {Bernu}, \citenamefont {Del\'{e}glise},
  \citenamefont {Sayrin}, \citenamefont {Gleyzes}, \citenamefont {Kuhr},
  \citenamefont {Brune}, \citenamefont {Raimond},\ and\ \citenamefont
  {Haroche}}]{Collapse}%
  \BibitemOpen
  \bibfield  {author} {\bibinfo {author} {\bibfnamefont {C.}~\bibnamefont
  {Guerlin}}, \bibinfo {author} {\bibfnamefont {J.}~\bibnamefont {Bernu}},
  \bibinfo {author} {\bibfnamefont {S.}~\bibnamefont {Del\'{e}glise}}, \bibinfo
  {author} {\bibfnamefont {C.}~\bibnamefont {Sayrin}}, \bibinfo {author}
  {\bibfnamefont {S.}~\bibnamefont {Gleyzes}}, \bibinfo {author} {\bibfnamefont
  {S.}~\bibnamefont {Kuhr}}, \bibinfo {author} {\bibfnamefont {M.}~\bibnamefont
  {Brune}}, \bibinfo {author} {\bibfnamefont {J.}~\bibnamefont {Raimond}}, \
  and\ \bibinfo {author} {\bibfnamefont {S.}~\bibnamefont {Haroche}},\
  }\href@noop {} {\bibfield  {journal} {\bibinfo  {journal} {Nature}\ }\textbf
  {\bibinfo {volume} {448}},\ \bibinfo {pages} {889} (\bibinfo {year}
  {2007})}\BibitemShut {NoStop}%
\bibitem [{\citenamefont {Nogues}\ \emph {et~al.}(1999)\citenamefont {Nogues},
  \citenamefont {Rauschenbeutel}, \citenamefont {Osnaghi}, \citenamefont
  {Brune}, \citenamefont {Raimond},\ and\ \citenamefont {Haroche}}]{ROQND}%
  \BibitemOpen
  \bibfield  {author} {\bibinfo {author} {\bibfnamefont {G.}~\bibnamefont
  {Nogues}}, \bibinfo {author} {\bibfnamefont {A.}~\bibnamefont
  {Rauschenbeutel}}, \bibinfo {author} {\bibfnamefont {S.}~\bibnamefont
  {Osnaghi}}, \bibinfo {author} {\bibfnamefont {M.}~\bibnamefont {Brune}},
  \bibinfo {author} {\bibfnamefont {J.~M.}\ \bibnamefont {Raimond}}, \ and\
  \bibinfo {author} {\bibfnamefont {S.}~\bibnamefont {Haroche}},\ }\href@noop
  {} {\bibfield  {journal} {\bibinfo  {journal} {Nature}\ }\textbf {\bibinfo
  {volume} {400}},\ \bibinfo {pages} {239} (\bibinfo {year}
  {1999})}\BibitemShut {NoStop}%
\bibitem [{\citenamefont {Johnson}\ \emph {et~al.}(2010)\citenamefont
  {Johnson}, \citenamefont {Reed}, \citenamefont {Houck}, \citenamefont
  {Schuster}, \citenamefont {Bishop}, \citenamefont {Ginossar}, \citenamefont
  {Gambetta}, \citenamefont {DiCarlo}, \citenamefont {Frunzio}, \citenamefont
  {Girvin},\ and\ \citenamefont {Schoelkopf}}]{QGate}%
  \BibitemOpen
  \bibfield  {author} {\bibinfo {author} {\bibfnamefont {B.~R.}\ \bibnamefont
  {Johnson}}, \bibinfo {author} {\bibfnamefont {M.~D.}\ \bibnamefont {Reed}},
  \bibinfo {author} {\bibfnamefont {A.~A.}\ \bibnamefont {Houck}}, \bibinfo
  {author} {\bibfnamefont {D.~I.}\ \bibnamefont {Schuster}}, \bibinfo {author}
  {\bibfnamefont {L.~S.}\ \bibnamefont {Bishop}}, \bibinfo {author}
  {\bibfnamefont {E.}~\bibnamefont {Ginossar}}, \bibinfo {author}
  {\bibfnamefont {J.~M.}\ \bibnamefont {Gambetta}}, \bibinfo {author}
  {\bibfnamefont {L.}~\bibnamefont {DiCarlo}}, \bibinfo {author} {\bibfnamefont
  {L.}~\bibnamefont {Frunzio}}, \bibinfo {author} {\bibfnamefont {S.~M.}\
  \bibnamefont {Girvin}}, \ and\ \bibinfo {author} {\bibfnamefont {R.~J.}\
  \bibnamefont {Schoelkopf}},\ }\href@noop {} {\bibfield  {journal} {\bibinfo
  {journal} {Nat. Phys.}\ }\textbf {\bibinfo {volume} {6}},\ \bibinfo {pages}
  {663} (\bibinfo {year} {2010})}\BibitemShut {NoStop}%
\bibitem [{\citenamefont {Brune}\ \emph {et~al.}(1990)\citenamefont {Brune},
  \citenamefont {Haroche}, \citenamefont {Lefevre}, \citenamefont {Raimond},\
  and\ \citenamefont {Zagury}}]{Stark1}%
  \BibitemOpen
  \bibfield  {author} {\bibinfo {author} {\bibfnamefont {M.}~\bibnamefont
  {Brune}}, \bibinfo {author} {\bibfnamefont {S.}~\bibnamefont {Haroche}},
  \bibinfo {author} {\bibfnamefont {V.}~\bibnamefont {Lefevre}}, \bibinfo
  {author} {\bibfnamefont {J.~M.}\ \bibnamefont {Raimond}}, \ and\ \bibinfo
  {author} {\bibfnamefont {N.}~\bibnamefont {Zagury}},\ }\href@noop {}
  {\bibfield  {journal} {\bibinfo  {journal} {Phys. Rev. Lett.}\ }\textbf
  {\bibinfo {volume} {65}},\ \bibinfo {pages} {976} (\bibinfo {year}
  {1990})}\BibitemShut {NoStop}%
\bibitem [{\citenamefont {Schuster}\ \emph {et~al.}(2007)\citenamefont
  {Schuster}, \citenamefont {Houck}, \citenamefont {Schreier1}, \citenamefont
  {Wallraff}, \citenamefont {Gambetta}, \citenamefont {Blais}, \citenamefont
  {Frunzio}, \citenamefont {Majer}, \citenamefont {Johnson}, \citenamefont
  {Devoret}, \citenamefont {Girvin},\ and\ \citenamefont
  {Schoelkopf}}]{Stark2}%
  \BibitemOpen
  \bibfield  {author} {\bibinfo {author} {\bibfnamefont {D.~I.}\ \bibnamefont
  {Schuster}}, \bibinfo {author} {\bibfnamefont {A.~A.}\ \bibnamefont {Houck}},
  \bibinfo {author} {\bibfnamefont {J.~A.}\ \bibnamefont {Schreier1}}, \bibinfo
  {author} {\bibfnamefont {A.}~\bibnamefont {Wallraff}}, \bibinfo {author}
  {\bibfnamefont {J.~M.}\ \bibnamefont {Gambetta}}, \bibinfo {author}
  {\bibfnamefont {A.}~\bibnamefont {Blais}}, \bibinfo {author} {\bibfnamefont
  {L.}~\bibnamefont {Frunzio}}, \bibinfo {author} {\bibfnamefont
  {J.}~\bibnamefont {Majer}}, \bibinfo {author} {\bibfnamefont
  {B.}~\bibnamefont {Johnson}}, \bibinfo {author} {\bibfnamefont {M.~H.}\
  \bibnamefont {Devoret}}, \bibinfo {author} {\bibfnamefont {S.~M.}\
  \bibnamefont {Girvin}}, \ and\ \bibinfo {author} {\bibfnamefont {R.~J.}\
  \bibnamefont {Schoelkopf}},\ }\href@noop {} {\bibfield  {journal} {\bibinfo
  {journal} {Nature}\ }\textbf {\bibinfo {volume} {445}},\ \bibinfo {pages}
  {515} (\bibinfo {year} {2007})}\BibitemShut {NoStop}%
\bibitem [{\citenamefont {Reiserer}\ \emph {et~al.}(2013)\citenamefont
  {Reiserer}, \citenamefont {Ritter},\ and\ \citenamefont
  {Rempe}}]{QNDOpthicalPhoton}%
  \BibitemOpen
  \bibfield  {author} {\bibinfo {author} {\bibfnamefont {A.}~\bibnamefont
  {Reiserer}}, \bibinfo {author} {\bibfnamefont {S.}~\bibnamefont {Ritter}}, \
  and\ \bibinfo {author} {\bibfnamefont {G.}~\bibnamefont {Rempe}},\
  }\href@noop {} {\bibfield  {journal} {\bibinfo  {journal} {Science}\ }\textbf
  {\bibinfo {volume} {342}},\ \bibinfo {pages} {1349} (\bibinfo {year}
  {2013})}\BibitemShut {NoStop}%
\bibitem [{\citenamefont {Sathyamoorthy}\ \emph {et~al.}(2014)\citenamefont
  {Sathyamoorthy}, \citenamefont {Tornberg}, \citenamefont {Kockum},
  \citenamefont {Baragiola}, \citenamefont {Combes}, \citenamefont {Wilson},
  \citenamefont {Stace},\ and\ \citenamefont {Johansson}}]{QNDTravelMW}%
  \BibitemOpen
  \bibfield  {author} {\bibinfo {author} {\bibfnamefont {S.~R.}\ \bibnamefont
  {Sathyamoorthy}}, \bibinfo {author} {\bibfnamefont {L.}~\bibnamefont
  {Tornberg}}, \bibinfo {author} {\bibfnamefont {A.~F.}\ \bibnamefont
  {Kockum}}, \bibinfo {author} {\bibfnamefont {B.~Q.}\ \bibnamefont
  {Baragiola}}, \bibinfo {author} {\bibfnamefont {J.}~\bibnamefont {Combes}},
  \bibinfo {author} {\bibfnamefont {C.~M.}\ \bibnamefont {Wilson}}, \bibinfo
  {author} {\bibfnamefont {T.~M.}\ \bibnamefont {Stace}}, \ and\ \bibinfo
  {author} {\bibfnamefont {G.}~\bibnamefont {Johansson}},\ }\href@noop {}
  {\bibfield  {journal} {\bibinfo  {journal} {Phys. Rev. Lett.}\ }\textbf
  {\bibinfo {volume} {112}},\ \bibinfo {pages} {093601} (\bibinfo {year}
  {2014})}\BibitemShut {NoStop}%
\bibitem [{\citenamefont {Milburn}(1989)}]{MilburnXPM}%
  \BibitemOpen
  \bibfield  {author} {\bibinfo {author} {\bibfnamefont {G.~J.}\ \bibnamefont
  {Milburn}},\ }\href@noop {} {\bibfield  {journal} {\bibinfo  {journal} {Phys.
  Rev. Lett.}\ }\textbf {\bibinfo {volume} {62}},\ \bibinfo {pages} {2124}
  (\bibinfo {year} {1989})}\BibitemShut {NoStop}%
\bibitem [{\citenamefont {Li}\ \emph {et~al.}(2013)\citenamefont {Li},
  \citenamefont {Deng},\ and\ \citenamefont {Hagley}}]{XPMExp1}%
  \BibitemOpen
  \bibfield  {author} {\bibinfo {author} {\bibfnamefont {R.~B.}\ \bibnamefont
  {Li}}, \bibinfo {author} {\bibfnamefont {L.}~\bibnamefont {Deng}}, \ and\
  \bibinfo {author} {\bibfnamefont {E.~W.}\ \bibnamefont {Hagley}},\
  }\href@noop {} {\bibfield  {journal} {\bibinfo  {journal} {Phys. Rev. Lett.}\
  }\textbf {\bibinfo {volume} {110}},\ \bibinfo {pages} {113902} (\bibinfo
  {year} {2013})}\BibitemShut {NoStop}%
\bibitem [{\citenamefont {Grangier}\ \emph {et~al.}(1998)\citenamefont
  {Grangier}, \citenamefont {Levenson},\ and\ \citenamefont
  {Poizat}}]{QNDReview}%
  \BibitemOpen
  \bibfield  {author} {\bibinfo {author} {\bibfnamefont {P.}~\bibnamefont
  {Grangier}}, \bibinfo {author} {\bibfnamefont {J.~A.}\ \bibnamefont
  {Levenson}}, \ and\ \bibinfo {author} {\bibfnamefont {J.}~\bibnamefont
  {Poizat}},\ }\href@noop {} {\bibfield  {journal} {\bibinfo  {journal}
  {Nature}\ }\textbf {\bibinfo {volume} {396}},\ \bibinfo {pages} {537}
  (\bibinfo {year} {1998})}\BibitemShut {NoStop}%
\bibitem [{\citenamefont {Lukin}\ and\ \citenamefont
  {Imamo\v{g}lu}(2000)}]{XPM}%
  \BibitemOpen
  \bibfield  {author} {\bibinfo {author} {\bibfnamefont {M.~D.}\ \bibnamefont
  {Lukin}}\ and\ \bibinfo {author} {\bibfnamefont {A.}~\bibnamefont
  {Imamo\v{g}lu}},\ }\href@noop {} {\bibfield  {journal} {\bibinfo  {journal}
  {Phys. Rev. Lett.}\ }\textbf {\bibinfo {volume} {84}},\ \bibinfo {pages}
  {1419} (\bibinfo {year} {2000})}\BibitemShut {NoStop}%
\bibitem [{\citenamefont {Fan}\ \emph {et~al.}(2013)\citenamefont {Fan},
  \citenamefont {Kockum}, \citenamefont {Combes}, \citenamefont {Johansson},
  \citenamefont {Hoi}, \citenamefont {Wilson}, \citenamefont {Delsing},
  \citenamefont {Milburn},\ and\ \citenamefont {Stace}}]{KerrDebate1}%
  \BibitemOpen
  \bibfield  {author} {\bibinfo {author} {\bibfnamefont {B.}~\bibnamefont
  {Fan}}, \bibinfo {author} {\bibfnamefont {A.~F.}\ \bibnamefont {Kockum}},
  \bibinfo {author} {\bibfnamefont {J.}~\bibnamefont {Combes}}, \bibinfo
  {author} {\bibfnamefont {G.}~\bibnamefont {Johansson}}, \bibinfo {author}
  {\bibfnamefont {I.-c.}\ \bibnamefont {Hoi}}, \bibinfo {author} {\bibfnamefont
  {C.~M.}\ \bibnamefont {Wilson}}, \bibinfo {author} {\bibfnamefont
  {P.}~\bibnamefont {Delsing}}, \bibinfo {author} {\bibfnamefont {G.~J.}\
  \bibnamefont {Milburn}}, \ and\ \bibinfo {author} {\bibfnamefont {T.~M.}\
  \bibnamefont {Stace}},\ }\href@noop {} {\bibfield  {journal} {\bibinfo
  {journal} {Phys. Rev. Lett.}\ }\textbf {\bibinfo {volume} {110}},\ \bibinfo
  {pages} {053601} (\bibinfo {year} {2013})}\BibitemShut {NoStop}%
\bibitem [{\citenamefont {Gea-Banacloche}(2010)}]{KerrDebate2}%
  \BibitemOpen
  \bibfield  {author} {\bibinfo {author} {\bibfnamefont {J.}~\bibnamefont
  {Gea-Banacloche}},\ }\href@noop {} {\bibfield  {journal} {\bibinfo  {journal}
  {Phys. Rev. A}\ }\textbf {\bibinfo {volume} {81}},\ \bibinfo {pages} {043823}
  (\bibinfo {year} {2010})}\BibitemShut {NoStop}%
\bibitem [{\citenamefont {Shapiro}\ and\ \citenamefont
  {Razavi}(2007)}]{KerrDebate4}%
  \BibitemOpen
  \bibfield  {author} {\bibinfo {author} {\bibfnamefont {J.~H.}\ \bibnamefont
  {Shapiro}}\ and\ \bibinfo {author} {\bibfnamefont {M.}~\bibnamefont
  {Razavi}},\ }\href@noop {} {\bibfield  {journal} {\bibinfo  {journal} {New J.
  Phys.}\ }\textbf {\bibinfo {volume} {9}},\ \bibinfo {pages} {16} (\bibinfo
  {year} {2007})}\BibitemShut {NoStop}%
\bibitem [{\citenamefont {Shapiro}(2006)}]{KerrDebate3}%
  \BibitemOpen
  \bibfield  {author} {\bibinfo {author} {\bibfnamefont {J.~H.}\ \bibnamefont
  {Shapiro}},\ }\href@noop {} {\bibfield  {journal} {\bibinfo  {journal} {Phys.
  Rev. A}\ }\textbf {\bibinfo {volume} {73}},\ \bibinfo {pages} {062305}
  (\bibinfo {year} {2006})}\BibitemShut {NoStop}%
\bibitem [{\citenamefont {Dove}\ \emph
  {et~al.}(2014{\natexlab{a}})\citenamefont {Dove}, \citenamefont {Chudzicki},\
  and\ \citenamefont {Shapiro}}]{ShapiroXPM}%
  \BibitemOpen
  \bibfield  {author} {\bibinfo {author} {\bibfnamefont {J.}~\bibnamefont
  {Dove}}, \bibinfo {author} {\bibfnamefont {C.}~\bibnamefont {Chudzicki}}, \
  and\ \bibinfo {author} {\bibfnamefont {J.~H.}\ \bibnamefont {Shapiro}},\
  }\href@noop {} {\bibfield  {journal} {\bibinfo  {journal} {Phys. Rev. A}\
  }\textbf {\bibinfo {volume} {90}},\ \bibinfo {pages} {062314} (\bibinfo
  {year} {2014}{\natexlab{a}})}\BibitemShut {NoStop}%
\bibitem [{\citenamefont {Langford}\ \emph {et~al.}(2011)\citenamefont
  {Langford}, \citenamefont {Ramelow}, \citenamefont {Prevedel}, \citenamefont
  {Munro}, \citenamefont {Milburn},\ and\ \citenamefont
  {Zeilinger}}]{Nature478p360}%
  \BibitemOpen
  \bibfield  {author} {\bibinfo {author} {\bibfnamefont {N.~K.}\ \bibnamefont
  {Langford}}, \bibinfo {author} {\bibfnamefont {S.}~\bibnamefont {Ramelow}},
  \bibinfo {author} {\bibfnamefont {R.}~\bibnamefont {Prevedel}}, \bibinfo
  {author} {\bibfnamefont {W.~J.}\ \bibnamefont {Munro}}, \bibinfo {author}
  {\bibfnamefont {G.~J.}\ \bibnamefont {Milburn}}, \ and\ \bibinfo {author}
  {\bibfnamefont {A.}~\bibnamefont {Zeilinger}},\ }\href@noop {} {\bibfield
  {journal} {\bibinfo  {journal} {Nature}\ }\textbf {\bibinfo {volume} {478}},\
  \bibinfo {pages} {360} (\bibinfo {year} {2011})}\BibitemShut {NoStop}%
\bibitem [{\citenamefont {Bertet}\ \emph {et~al.}(2002)\citenamefont {Bertet},
  \citenamefont {Auffeves}, \citenamefont {Maioli}, \citenamefont {Osnaghi},
  \citenamefont {Meunier}, \citenamefont {Brune}, \citenamefont {Raimond},\
  and\ \citenamefont {Haroche}}]{HarocheDispl}%
  \BibitemOpen
  \bibfield  {author} {\bibinfo {author} {\bibfnamefont {P.}~\bibnamefont
  {Bertet}}, \bibinfo {author} {\bibfnamefont {A.}~\bibnamefont {Auffeves}},
  \bibinfo {author} {\bibfnamefont {P.}~\bibnamefont {Maioli}}, \bibinfo
  {author} {\bibfnamefont {S.}~\bibnamefont {Osnaghi}}, \bibinfo {author}
  {\bibfnamefont {T.}~\bibnamefont {Meunier}}, \bibinfo {author} {\bibfnamefont
  {M.}~\bibnamefont {Brune}}, \bibinfo {author} {\bibfnamefont {J.~M.}\
  \bibnamefont {Raimond}}, \ and\ \bibinfo {author} {\bibfnamefont
  {S.}~\bibnamefont {Haroche}},\ }\href@noop {} {\bibfield  {journal} {\bibinfo
   {journal} {Phys. Rev. Lett.}\ }\textbf {\bibinfo {volume} {89}},\ \bibinfo
  {pages} {200402} (\bibinfo {year} {2002})}\BibitemShut {NoStop}%
\bibitem [{\citenamefont {Banaszek}\ and\ \citenamefont
  {W\'odkiewicz}(1996)}]{BS1}%
  \BibitemOpen
  \bibfield  {author} {\bibinfo {author} {\bibfnamefont {K.}~\bibnamefont
  {Banaszek}}\ and\ \bibinfo {author} {\bibfnamefont {K.}~\bibnamefont
  {W\'odkiewicz}},\ }\href@noop {} {\bibfield  {journal} {\bibinfo  {journal}
  {Phys. Rev. Lett.}\ }\textbf {\bibinfo {volume} {76}},\ \bibinfo {pages}
  {4344} (\bibinfo {year} {1996})}\BibitemShut {NoStop}%
\bibitem [{\citenamefont {Barnett}\ and\ \citenamefont
  {Croke}(2009)}]{StatePerr}%
  \BibitemOpen
  \bibfield  {author} {\bibinfo {author} {\bibfnamefont {S.~M.}\ \bibnamefont
  {Barnett}}\ and\ \bibinfo {author} {\bibfnamefont {S.}~\bibnamefont
  {Croke}},\ }\href@noop {} {\bibfield  {journal} {\bibinfo  {journal} {Adv.
  Opt. Photon.}\ }\textbf {\bibinfo {volume} {1}},\ \bibinfo {pages} {238}
  (\bibinfo {year} {2009})}\BibitemShut {NoStop}%
\bibitem [{\citenamefont {Brune}\ \emph {et~al.}(1996)\citenamefont {Brune},
  \citenamefont {Schmidt-Kaler}, \citenamefont {Maali}, \citenamefont {Dreyer},
  \citenamefont {Hagley}, \citenamefont {Raimond},\ and\ \citenamefont
  {Haroche}}]{FockDispersion}%
  \BibitemOpen
  \bibfield  {author} {\bibinfo {author} {\bibfnamefont {M.}~\bibnamefont
  {Brune}}, \bibinfo {author} {\bibfnamefont {F.}~\bibnamefont
  {Schmidt-Kaler}}, \bibinfo {author} {\bibfnamefont {A.}~\bibnamefont
  {Maali}}, \bibinfo {author} {\bibfnamefont {J.}~\bibnamefont {Dreyer}},
  \bibinfo {author} {\bibfnamefont {E.}~\bibnamefont {Hagley}}, \bibinfo
  {author} {\bibfnamefont {J.~M.}\ \bibnamefont {Raimond}}, \ and\ \bibinfo
  {author} {\bibfnamefont {S.}~\bibnamefont {Haroche}},\ }\href@noop {}
  {\bibfield  {journal} {\bibinfo  {journal} {Phys. Rev. Lett.}\ }\textbf
  {\bibinfo {volume} {76}},\ \bibinfo {pages} {1800} (\bibinfo {year}
  {1996})}\BibitemShut {NoStop}%
\bibitem [{\citenamefont {Shen}\ \emph {et~al.}(2011)\citenamefont {Shen},
  \citenamefont {Bradford},\ and\ \citenamefont
  {Shen}}]{SinglePhotonRealSpace1}%
  \BibitemOpen
  \bibfield  {author} {\bibinfo {author} {\bibfnamefont {Y.}~\bibnamefont
  {Shen}}, \bibinfo {author} {\bibfnamefont {M.}~\bibnamefont {Bradford}}, \
  and\ \bibinfo {author} {\bibfnamefont {J.-T.}\ \bibnamefont {Shen}},\
  }\href@noop {} {\bibfield  {journal} {\bibinfo  {journal} {Phys. Rev. Lett.}\
  }\textbf {\bibinfo {volume} {107}},\ \bibinfo {pages} {173902} (\bibinfo
  {year} {2011})}\BibitemShut {NoStop}%
\bibitem [{\citenamefont {Bradford}\ and\ \citenamefont
  {Shen}(2012)}]{SinglePhotonRealSpace5}%
  \BibitemOpen
  \bibfield  {author} {\bibinfo {author} {\bibfnamefont {M.}~\bibnamefont
  {Bradford}}\ and\ \bibinfo {author} {\bibfnamefont {J.~T.}\ \bibnamefont
  {Shen}},\ }\href@noop {} {\bibfield  {journal} {\bibinfo  {journal} {Phys.
  Rev. A}\ }\textbf {\bibinfo {volume} {85}},\ \bibinfo {pages} {043814}
  (\bibinfo {year} {2012})}\BibitemShut {NoStop}%
\bibitem [{\citenamefont {Shen}\ and\ \citenamefont
  {Fan}(2009)}]{SinglePhotonRealSpace6}%
  \BibitemOpen
  \bibfield  {author} {\bibinfo {author} {\bibfnamefont {J.~T.}\ \bibnamefont
  {Shen}}\ and\ \bibinfo {author} {\bibfnamefont {S.}~\bibnamefont {Fan}},\
  }\href@noop {} {\bibfield  {journal} {\bibinfo  {journal} {Phys. Rev. A}\
  }\textbf {\bibinfo {volume} {79}},\ \bibinfo {pages} {023837} (\bibinfo
  {year} {2009})}\BibitemShut {NoStop}%
\bibitem [{\citenamefont {Xia}(2014)}]{SinglePhotonRealSpace2}%
  \BibitemOpen
  \bibfield  {author} {\bibinfo {author} {\bibfnamefont {K.}~\bibnamefont
  {Xia}},\ }\href@noop {} {\bibfield  {journal} {\bibinfo  {journal} {Phys.
  Rev. A}\ }\textbf {\bibinfo {volume} {89}},\ \bibinfo {pages} {023815}
  (\bibinfo {year} {2014})}\BibitemShut {NoStop}%
\bibitem [{\citenamefont {Xia}\ \emph {et~al.}(2014)\citenamefont {Xia},
  \citenamefont {Lu}, \citenamefont {Lin}, \citenamefont {Cheng}, \citenamefont
  {Niu}, \citenamefont {Gong},\ and\ \citenamefont
  {Twamley}}]{SinglePhotonRealSpace3}%
  \BibitemOpen
  \bibfield  {author} {\bibinfo {author} {\bibfnamefont {K.}~\bibnamefont
  {Xia}}, \bibinfo {author} {\bibfnamefont {G.}~\bibnamefont {Lu}}, \bibinfo
  {author} {\bibfnamefont {G.}~\bibnamefont {Lin}}, \bibinfo {author}
  {\bibfnamefont {Y.}~\bibnamefont {Cheng}}, \bibinfo {author} {\bibfnamefont
  {Y.}~\bibnamefont {Niu}}, \bibinfo {author} {\bibfnamefont {S.}~\bibnamefont
  {Gong}}, \ and\ \bibinfo {author} {\bibfnamefont {J.}~\bibnamefont
  {Twamley}},\ }\href@noop {} {\bibfield  {journal} {\bibinfo  {journal} {Phys.
  Rev. A}\ }\textbf {\bibinfo {volume} {90}},\ \bibinfo {pages} {043802}
  (\bibinfo {year} {2014})}\BibitemShut {NoStop}%
\bibitem [{\citenamefont {III}\ \emph {et~al.}(2010)\citenamefont {III},
  \citenamefont {Elshaari},\ and\ \citenamefont
  {Preble}}]{SinglePhotonRealSpace4}%
  \BibitemOpen
  \bibfield  {author} {\bibinfo {author} {\bibfnamefont {E.~E.~H.}\
  \bibnamefont {III}}, \bibinfo {author} {\bibfnamefont {A.~W.}\ \bibnamefont
  {Elshaari}}, \ and\ \bibinfo {author} {\bibfnamefont {S.~F.}\ \bibnamefont
  {Preble}},\ }\href@noop {} {\bibfield  {journal} {\bibinfo  {journal} {Phys.
  Rev. A}\ }\textbf {\bibinfo {volume} {82}},\ \bibinfo {pages} {063839}
  (\bibinfo {year} {2010})}\BibitemShut {NoStop}%
\bibitem [{\citenamefont {Hilfer}\ and\ \citenamefont
  {Menyuk}(1992)}]{NonlocalKerr1}%
  \BibitemOpen
  \bibfield  {author} {\bibinfo {author} {\bibfnamefont {G.}~\bibnamefont
  {Hilfer}}\ and\ \bibinfo {author} {\bibfnamefont {C.~R.}\ \bibnamefont
  {Menyuk}},\ }\href@noop {} {\bibfield  {journal} {\bibinfo  {journal} {Opt.
  Lett.}\ }\textbf {\bibinfo {volume} {17}},\ \bibinfo {pages} {949} (\bibinfo
  {year} {1992})}\BibitemShut {NoStop}%
\bibitem [{\citenamefont {Pepper}\ \emph {et~al.}(1978)\citenamefont {Pepper},
  \citenamefont {AuYeung}, \citenamefont {Fekete},\ and\ \citenamefont
  {Yariv}}]{NonlocalKerr2}%
  \BibitemOpen
  \bibfield  {author} {\bibinfo {author} {\bibfnamefont {D.~M.}\ \bibnamefont
  {Pepper}}, \bibinfo {author} {\bibfnamefont {J.}~\bibnamefont {AuYeung}},
  \bibinfo {author} {\bibfnamefont {D.}~\bibnamefont {Fekete}}, \ and\ \bibinfo
  {author} {\bibfnamefont {A.}~\bibnamefont {Yariv}},\ }\href@noop {}
  {\bibfield  {journal} {\bibinfo  {journal} {Opt. Lett.}\ }\textbf {\bibinfo
  {volume} {3}},\ \bibinfo {pages} {7} (\bibinfo {year} {1978})}\BibitemShut
  {NoStop}%
\bibitem [{\citenamefont {Wang}\ \emph {et~al.}(2012)\citenamefont {Wang},
  \citenamefont {Zheng}, \citenamefont {Zhou}, \citenamefont {Yang},
  \citenamefont {Lu}, \citenamefont {Guo},\ and\ \citenamefont
  {Hu}}]{NonlocalKerr3}%
  \BibitemOpen
  \bibfield  {author} {\bibinfo {author} {\bibfnamefont {J.}~\bibnamefont
  {Wang}}, \bibinfo {author} {\bibfnamefont {Y.-Z.}\ \bibnamefont {Zheng}},
  \bibinfo {author} {\bibfnamefont {L.-H.}\ \bibnamefont {Zhou}}, \bibinfo
  {author} {\bibfnamefont {Z.-J.}\ \bibnamefont {Yang}}, \bibinfo {author}
  {\bibfnamefont {D.-Q.}\ \bibnamefont {Lu}}, \bibinfo {author} {\bibfnamefont
  {Q.}~\bibnamefont {Guo}}, \ and\ \bibinfo {author} {\bibfnamefont
  {W.}~\bibnamefont {Hu}},\ }\href@noop {} {\bibfield  {journal} {\bibinfo
  {journal} {Acta Phys. Sin.}\ }\textbf {\bibinfo {volume} {61}},\ \bibinfo
  {pages} {084210} (\bibinfo {year} {2012})}\BibitemShut {NoStop}%
\bibitem [{\citenamefont {Engin}\ \emph {et~al.}(1997)\citenamefont {Engin},
  \citenamefont {Cross},\ and\ \citenamefont {Yariv}}]{NonlocalKerr4}%
  \BibitemOpen
  \bibfield  {author} {\bibinfo {author} {\bibfnamefont {D.}~\bibnamefont
  {Engin}}, \bibinfo {author} {\bibfnamefont {M.~C.}\ \bibnamefont {Cross}}, \
  and\ \bibinfo {author} {\bibfnamefont {A.}~\bibnamefont {Yariv}},\
  }\href@noop {} {\bibfield  {journal} {\bibinfo  {journal} {J. Opt. Soc. Am.
  B}\ }\textbf {\bibinfo {volume} {14}},\ \bibinfo {pages} {3349} (\bibinfo
  {year} {1997})}\BibitemShut {NoStop}%
\bibitem [{\citenamefont {Biaggio}(1999)}]{NonlocalKerr5}%
  \BibitemOpen
  \bibfield  {author} {\bibinfo {author} {\bibfnamefont {I.}~\bibnamefont
  {Biaggio}},\ }\href@noop {} {\bibfield  {journal} {\bibinfo  {journal} {Phys.
  Rev. Lett.}\ }\textbf {\bibinfo {volume} {82}},\ \bibinfo {pages} {193}
  (\bibinfo {year} {1999})}\BibitemShut {NoStop}%
\bibitem [{\citenamefont {Bar-Ad}\ \emph {et~al.}(2013)\citenamefont {Bar-Ad},
  \citenamefont {Schilling},\ and\ \citenamefont {Fleurov}}]{NonlocalKerr6}%
  \BibitemOpen
  \bibfield  {author} {\bibinfo {author} {\bibfnamefont {S.}~\bibnamefont
  {Bar-Ad}}, \bibinfo {author} {\bibfnamefont {R.}~\bibnamefont {Schilling}}, \
  and\ \bibinfo {author} {\bibfnamefont {V.}~\bibnamefont {Fleurov}},\
  }\href@noop {} {\bibfield  {journal} {\bibinfo  {journal} {arXiv:1301.3725}\
  } (\bibinfo {year} {2013})}\BibitemShut {NoStop}%
\bibitem [{\citenamefont {Dennis}\ \emph {et~al.}(2013)\citenamefont {Dennis},
  \citenamefont {Hope},\ and\ \citenamefont {Johnsson}}]{xmds}%
  \BibitemOpen
  \bibfield  {author} {\bibinfo {author} {\bibfnamefont {G.~R.}\ \bibnamefont
  {Dennis}}, \bibinfo {author} {\bibfnamefont {J.~J.}\ \bibnamefont {Hope}}, \
  and\ \bibinfo {author} {\bibfnamefont {M.~T.}\ \bibnamefont {Johnsson}},\
  }\href@noop {} {\bibfield  {journal} {\bibinfo  {journal} {Comp. Phys.
  Commun.}\ }\textbf {\bibinfo {volume} {184}},\ \bibinfo {pages} {201}
  (\bibinfo {year} {2013})}\BibitemShut {NoStop}%
\bibitem [{\citenamefont {Dove}\ \emph
  {et~al.}(2014{\natexlab{b}})\citenamefont {Dove}, \citenamefont {Chudzicki},\
  and\ \citenamefont {Shapiro}}]{ShapiroScanOver}%
  \BibitemOpen
  \bibfield  {author} {\bibinfo {author} {\bibfnamefont {J.}~\bibnamefont
  {Dove}}, \bibinfo {author} {\bibfnamefont {C.}~\bibnamefont {Chudzicki}}, \
  and\ \bibinfo {author} {\bibfnamefont {J.~H.}\ \bibnamefont {Shapiro}},\
  }\href {\doibase 10.1103/PhysRevA.90.062314} {\bibfield  {journal} {\bibinfo
  {journal} {Phys. Rev. A}\ }\textbf {\bibinfo {volume} {90}},\ \bibinfo
  {pages} {062314} (\bibinfo {year} {2014}{\natexlab{b}})}\BibitemShut
  {NoStop}%
\bibitem [{\citenamefont {Trita}\ \emph {et~al.}(2011)\citenamefont {Trita},
  \citenamefont {Lacava}, \citenamefont {Minzioni}, \citenamefont {Colonna},
  \citenamefont {Gautier}, \citenamefont {Fedeli},\ and\ \citenamefont
  {Cristiani}}]{trita2011}%
  \BibitemOpen
  \bibfield  {author} {\bibinfo {author} {\bibfnamefont {A.}~\bibnamefont
  {Trita}}, \bibinfo {author} {\bibfnamefont {C.}~\bibnamefont {Lacava}},
  \bibinfo {author} {\bibfnamefont {P.}~\bibnamefont {Minzioni}}, \bibinfo
  {author} {\bibfnamefont {J.-P.}\ \bibnamefont {Colonna}}, \bibinfo {author}
  {\bibfnamefont {P.}~\bibnamefont {Gautier}}, \bibinfo {author} {\bibfnamefont
  {J.-M.}\ \bibnamefont {Fedeli}}, \ and\ \bibinfo {author} {\bibfnamefont
  {I.}~\bibnamefont {Cristiani}},\ }\href@noop {} {\bibfield  {journal}
  {\bibinfo  {journal} {Appl. Phys. Lett.}\ }\textbf {\bibinfo {volume} {99}},\
  \bibinfo {pages} {191105} (\bibinfo {year} {2011})}\BibitemShut {NoStop}%
\end{thebibliography}
%

\end{document}